\documentclass[twocolumn,amsmath,showpacs,floatfix]{revtex4}
\usepackage{graphicx}

\newcommand{\degree}{\ensuremath{^{\circ}}}
\begin{document}

\title{New limit on Lorentz and CPT-violating neutron spin interactions}
\author{J. M. Brown}
\author{S. J. Smullin}
\author{T. W. Kornack}
\author{M. V. Romalis}

\begin{abstract}
We performed a search for neutron spin coupling to a Lorentz and
CPT-violating background field using a magnetometer with overlapping
ensembles of K and $^3$He atoms. The co-magnetometer is mounted on a
rotary platform for frequent reversal of its orientation. We measure
sidereal  oscillations in the signal to search for anomalous spin
coupling of extra-solar origin.  We determine the equatorial
components of the background field interacting with the neutron spin
to be $\widetilde{b}^n_X = (0.1 \pm 1.6) \times 10^{-33}$ GeV and
$\widetilde{b}^n_Y = (2.5 \pm 1.6) \times 10^{-33}$ GeV, improving
on the previous limit by a factor of 30. This measurement represents
the highest energy resolution of any spin anisotropy experiment.
\end{abstract}

\affiliation{Department of Physics, Princeton University, Princeton,
New Jersey 08544}


\maketitle

Experimental searches for anomalous spin coupling to an anisotropy
in space were first considered by Hughes \cite{Hughes} and Drever
\cite{Drever}. Since then, a number of such tests  have been
performed with electron  and nuclear spins with increasing
sensitivity
\cite{Prestage,Lamoreaux,Philips,Chupp,Hunter,Bear,Heckel}. There
has been a resurgence of interest in these searches following the
development of a general formalism for Lorentz and CPT violation
called the Standard Model Extension (SME) by Kosteleck$\acute{\rm
y}$ \cite{Kost}. The SME contains a number of possible terms that
violate local Lorentz invariance by coupling to particle spin
\cite{clockcom}. Furthermore, it has been shown that CPT violation
necessarily leads to Lorentz violation \cite{Greenberg}, opening the
possibility of CPT tests without the use of anti-particles.
Anisotropic spin coupling appears in a number of Lorentz-violating
models, for example, models with modified dispersion relationships
at high energy \cite{Posp1}, non-commutativity of space-time
\cite{Posp2}, and supersymmetric Lorentz violation \cite{Posp3}.
This suggests that it is a rather general feature of
Lorentz-violating theories.

The absolute energy sensitivity to anisotropic spin interactions is
a good figure of merit for Lorentz and CPT tests within the SME.
Previously, the most sensitive such test was performed for neutrons
using a $^3$He-$^{129}$Xe Zeeman maser \cite{Bear}. Here, we use a
K-$^3$He co-magnetometer to reach 0.7 nHz energy resolution,
improving the previous limit by a factor of 30. Existing limits on
possible electron interactions \cite{Heckel} and the simple nuclear
spin structure of $^3$He \cite{Friar} allow us to set clean limits
on nuclear spin Lorentz violation, mostly sensitive to neutron
interactions.

The K-$^3$He co-magnetometer is similar to that described in
\cite{gyro,spinforce} but is smaller in size. For this experiment
the entire optical setup is operated in vacuum to reduce low
frequency noise from air currents and the apparatus is mounted on a
rotary platform to reverse the direction of its sensitive axis every
22 seconds. We measure the sidereal oscillations of the signal to
remove Earth-fixed backgrounds, such as the gyroscopic signal due to
Earth's rotation.

\begin{figure}[tbp]
\centering
\includegraphics[width=8.5cm]{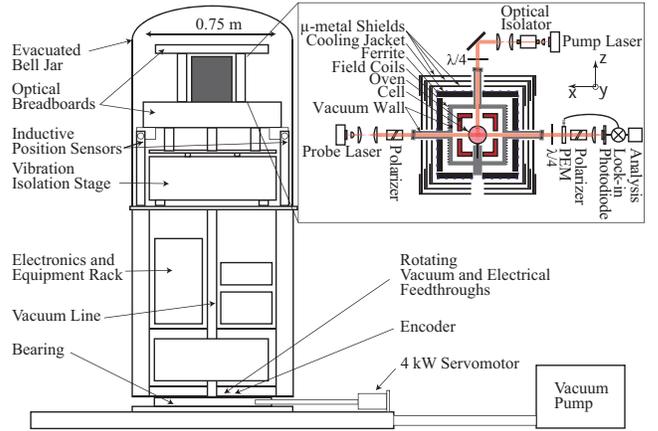}
\caption{Experimental setup of the rotating
co-magnetometer.}\label{fig_setup}
\end{figure}

The physics of the K-$^{3}$He co-magnetometer has been described
elsewhere \cite{se,gyro,Tomthesis}. Briefly, a circularly polarized
laser optically pumps a high density K vapor. Spin-exchange
collisions between K and $^{3}$He atoms polarize $^{3}$He spins.
These overlapping spin ensembles are coupled via spin-exchange and
their magnetic interaction.  An applied $B_z$ magnetic field
parallel to the pump beam cancels the effective magnetic field
experienced by K atoms due to nuclear spin magnetization of
$^{3}$He. As a result, the K magnetometer operates near zero  field,
where Zeeman resonance broadening due to spin-exchange collisions
between alkali-metal atoms is eliminated \cite{Hap}. At a particular
value of $B_z$ field called the compensation point the $\hat{x}$
polarization of K atoms has a particularly simple form,  given to
leading order by:
\begin{eqnarray}
P^{e}_x = \frac{P^{e}_{z}\gamma_{e}}{R_{tot}}
\left(\beta^{N}_{y}-\beta^{e}_{y}+\frac{\Omega_{y}}{\gamma_{N}}\right).
\label{signal}
\end{eqnarray}
Here, $\beta^{N}_{y}$ and $\beta^{e}_{y}$ describe the
phenomenological magnetic-like fields in the $\hat{y}$ direction
that couple only to $^{3}$He nucleus and K electrons respectively.
$P^{e}_{z}$  and $R_{tot}$ are the K electron spin polarization and
relaxation rate, $\gamma_{e}$  and $\gamma_{N}$ are the gyromagnetic
ratios for electrons and $^{3}$He nuclei respectively, and
$\Omega_y$ is the rotation rate of the apparatus.  Since K and
$^3$He atoms occupy the same volume, the co-magnetometer is
insensitive to ordinary magnetic fields
($\beta^{N}_{y}=\beta^{e}_{y}$), but retains sensitivity to
anomalous interactions that do not scale with the magnetic moment.

The experimental setup is shown in Figure \ref{fig_setup}. The atoms
are contained in a 2.4 cm diameter spherical cell made from
aluminosilicate glass filled with 9.4 amagats of $^3$He,  29 Torr of
N$_2$ for quenching, and a drop of K metal in the stem of the cell.
The cell is heated to 185\degree C by AC currents at 300 kHz in a
twisted pair wire heater, maintaining K density at $7\times
10^{13}$cm$^{-3}$.  A separate stem heater controls the position of
the K drop plugging the stem neck to preserve the spherical shape of
the polarized $^3$He. The magnetic shields consist of 3 layers of
$\mu$-metal and an inner ferrite shield to reduce thermal magnetic
noise and provide an overall shielding factor of 10$^8$
\cite{Ferrite}. Furthermore, a set of large Helmholtz coils surround
the apparatus to cancel the Earth's magnetic field and eliminate
Faraday rotation in optical elements. K atoms are optically pumped
in the $\hat{z}$ direction with about 10 mW of K D1 light generated
by a DFB laser. Coils inside the magnetic shields are used to cancel
residual magnetic fields and create the compensation $B_z$ field.
The polarization of K atoms in the $\hat{x}$ direction is measured
using optical rotation of a 10 mW linearly polarized off-resonant
probe beam generated by a DFB laser tuned to 770.76 nm. The residual
magnetic fields inside the shields and the lightshift due to probe
beam birefringence are eliminated using zeroing routines described
in \cite{gyro,Tomthesis}. The pump beam lightshift is reduced by
tuning the laser to the zero lightshift point on the D1 line and
monitoring its wavelength with a Burleigh WA-1500 wavemeter. The
volume around the cell is evacuated to 2 mTorr and  bell jar over
the entire optical setup is pumped out to 2 Torr to eliminate beam
motion due to air currents. We achieve sensitivity to
$\beta^{N}_{y}$, $\beta^{e}_{y}$ fields of 2 ${\rm fT}/\sqrt{{\rm
Hz}}$ at the apparatus reversal frequency of 0.023 Hz.

The optical setup and associated electronics are mounted on a rotary
platform driven through a worm-gear by an AC servo motor. Electric
power and vacuum connections are provided by rotary feedthroughs.
The experiment is controlled by a computer on the rotary platform
with a wireless internet connection. A rotary encoder measures the
angle of the platform and non-contact position sensors monitor the
orientation of the vibration isolation platform inside the bell jar.
The tilt of the rotation axis is measured with electronic tilt
sensors and zeroed to reduce laser beam motion correlated with
apparatus rotation.


\begin{figure}[tbp]
\centering
\includegraphics[width=8.5cm]{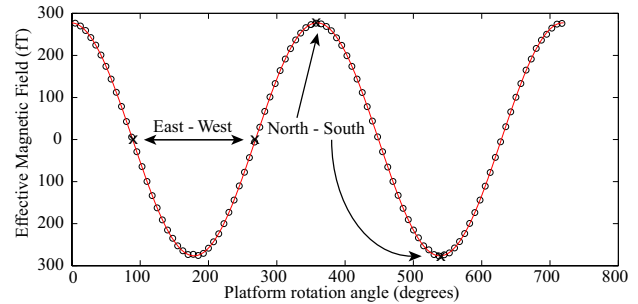}
\caption{Amplitude of the co-magnetometer signal for an 180\degree
rotation as a function of the initial platform rotation angle.  The
Lorentz violation data are collected at four points indicated by
crosses. }
\end{figure}

The co-magnetometer signal as a function of the rotation angle in
the lab frame is shown in Figure 2. A large signal is observed due
to the projection of the Earth's rotation onto the $\hat{y}$
direction. To remove this and other possible backgrounds fixed
relative to the Earth, we measure the sidereal oscillations in the
co-magnetometer signal amplitude resulting from 180$\degree$
reversal of the platform. We alternate either between North and
South orientations or East and West orientations every 22 sec. The
signal is recorded for 3-7 seconds while the apparatus is
stationary. Fast damping of spin transients in the co-magnetometer
\cite{gyro} are crucial for such data collection. After seven
orientation reversals, a 90$\degree$ rotation is performed to switch
to the other heading pair, the $B_z$ field is adjusted to the
compensation point and a sensitivity calibration is performed. From
these data, we extract the reversal-correlated amplitude.  Every 7
hours data collection pauses in the south position for zeroing of
other magnetic field components and probe light shift. These
operations are fully automated and the experiment can run for
several weeks without significant intervention. Figure 3 shows the
amplitude of the N-S and E-W signals as a function of the sidereal
time for a long run. The amplitude of the N-S signal agrees with the
projection of the Earth's rotation at our latitude within 2\%, and
the E-W signal is close to zero within the accuracy of the absolute
orientation of the co-magnetometer. In the presence of Lorentz
violation, each signal would exhibit a sinusoidal variation at the
sidereal frequency. We remove backgrounds slowly varying over
several days and fit the data to a sum of sine and cosine signals.
The error-bars of the fit amplitudes are increased by the reduced
$\chi^2$ of the fit, which is typically about 4.

The calibration of the co-magnetometer can be performed in several
ways. We usually use a calibration based on Eq. (5) of Ref.
\cite{gyro}, which can be performed at the same time as adjustment
of $B_z$ field. However, we found that it is sensitive to gradients
in magnetic field and alkali polarization. A more accurate
calibration was performed separately using slow modulation of the
$B_x$ field. A sinusoidal modulation  of $B_x$ field at  a frequency
$\omega$ with amplitude $B_0$ generates an out-of-phase response
given by
\begin{equation}
P^{e}_x=\frac{\gamma_eP_z^e}{R_{tot}}\frac{\omega B_0}{\gamma_N B_z}
\end{equation}
for $\omega \ll \gamma_N B_z$. It was verified using several other
calibration methods, such as rotation of the apparatus around the
vertical axis. Furthermore, the Earth's rotation rate provides an
additional check of the calibration.

\begin{figure}[tbp]
\centering
\includegraphics[width=8.5cm]{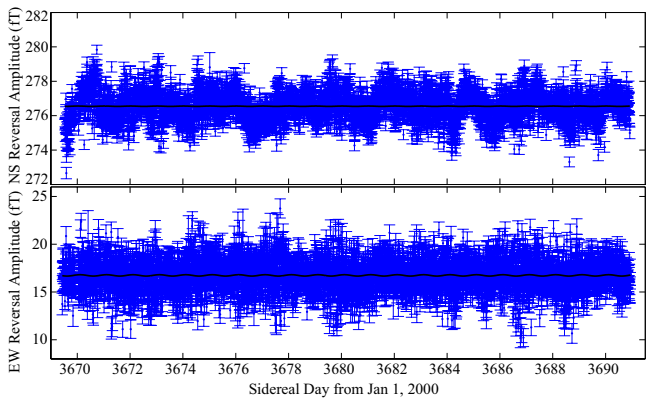}
\caption{Amplitude of the co-magnetometer signal for N-S and E-W
reversals as a function of sidereal time with a sinusoidal fit. Each
data point represents seven orientation reversals.  }
\end{figure}

The data were collected for 143 days from July 2009 to April 2010
and are shown in Figure 4. The long time span provides an important
separation between sidereal and possible diurnal variations. It is
also important to note that the N-S signal and the E-W signal are
susceptible to different systematic effects. The N-S signal is
mostly sensitive to changes in the co-magnetometer sensitivity,
while the E-W signal is sensitive to changes in the orientation of
the apparatus and drifts in the $B_z$ field and pump beam light
shift. Furthermore, a true sidereal signal would appear out-of-phase
in the N-S and E-W signals:
\begin{eqnarray}
S_{EW} &= &\beta_Y^N \cos(2 \pi t)-\beta_X^N \sin(2 \pi t) \\
S_{NS} &=&[-\beta_X^N \cos(2 \pi t)-\beta_Y^N \sin(2 \pi t)] \sin
\chi
\end{eqnarray}
where $\chi=40.35 \degree$ is the latitude at Princeton,
$\beta_X^N$, $\beta_Y^N$ are the components of the anomalous
magnetic field in the geocentric equatorial coordinate system
coupling to the $^3$He nuclear spin, and $t$ is the local sidereal
time.

A number of parameters of the experiment, such as various
temperatures, pump and probe beam positions, apparatus tilt, and
ambient magnetic fields were monitored, but no significant
correlations with the co-magnetometer signal were found. The only
finite correlation was due to drifts in the pump beam lightshift and
was corrected for based on laser wavelength measurements. To
estimate the systematic uncertainty, the data analysis was performed
using several methods. For example, the first point after pausing
for $B_z$ adjustment can be eliminated because of extra scatter
associated with vibration-isolation platform settling. The reversal
amplitude for each series of seven 180$\degree$ rotations can be
extrapolated to the time of $B_z$ adjustment to correct for drifts
of $^3$He polarization. The data can be fit separately for each
1-day interval to avoid potential bias from long-term drift removal.
All these methods and their combinations gave consistent results and
the scatter among them was used as a conservative estimate of the
systematic error.

\begin{figure}[tbp]
\centering
\includegraphics[width=8.5cm]{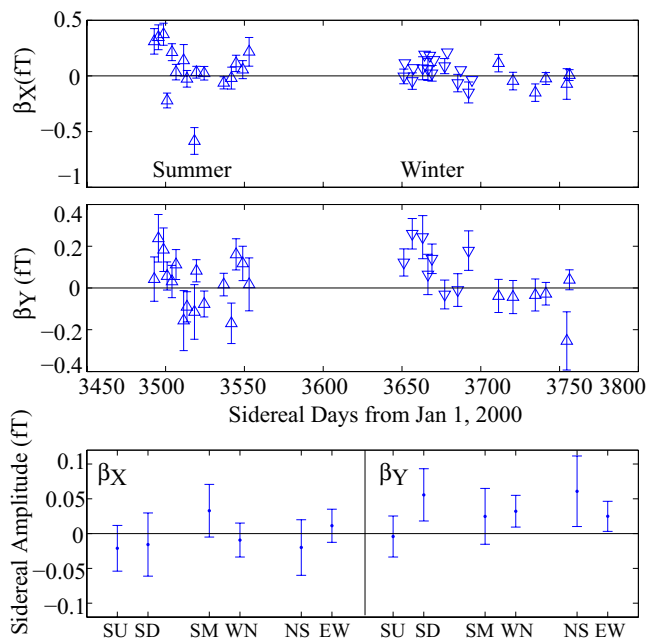}
\caption{Top panel: Summary of the data as a function of time. Each
data point represents a several-day run. Upward and downward
triangles represent direction of $^3$He spin. Bottom panel:
Systematic comparisons: SU/SD -spin up vs. spin down (from winter
data only), SM/WN - summer vs. winter data, EW/NS - amplitudes from
E-W vs. N-S reversals only.}
\end{figure}

It can be seen from Fig. 4 that the summer has a larger scatter of
data points than indicated by their error bars. We believe this is
due to fluctuations of the background optical rotation due to small
light interference effects affected by rotation of the apparatus. In
the winter, we implemented additional background measurements for
1.5-2 sec each time the apparatus was at rest by applying a large
$B_z$ field inside the shields to suppress rotation from the atoms.
Subtraction of this background from the signal reduces the scatter
of the data. In obtaining the final result, the summer and winter
data are averaged separately, scaling the errors by their respective
$\chi^2$ before the final average. The bottom panel of Fig. 4 shows
the comparison of the results taken in the summer vs. winter, with
the $^3$He spins up vs. down, and obtained separately from E-W and
N-S signals. It can be seen that all data subdivisions are
consistent with each other.

We obtain the following final result with statistical and systematic
errors:
\begin{eqnarray}
\beta_X^N &=& (0.001 \pm 0.019 \pm 0.010)~{\rm fT} \\
\beta_Y^N &=& (0.032 \pm 0.019 \pm 0.010)~{\rm fT}
\end{eqnarray}
In comparison, the limit on electron Lorentz violation from
\cite{Heckel} is equivalent to a magnetic field of 0.002 fT, so we
can ignore possible signal from electron interactions.

These results can be interpreted in terms of the parameters in the
SME \cite{Kost}. The following 3- and 4-dimensional operators in the
relativistic Lagrangian can be constrained from coupling to a
spin-1/2 particle
\begin{eqnarray}
\mathcal{L}&=&- \overline{\psi} (m+b_{\mu }\gamma ^{5}\gamma
^{\mu}+\frac{1}{2}
H_{\mu \nu }\sigma ^{\mu \nu })\psi \nonumber \\
&+&\frac{1}{2}i\overline{\psi}(\gamma _{\nu }+d_{\mu \nu }\gamma
^{5}\gamma ^{\mu }+\frac{1}{2} g_{\lambda \mu \nu} \sigma^{\lambda
\mu}){\overleftrightarrow \partial ^{\nu }}\psi. \label{Lagr}
\end{eqnarray}
Here $b_{\mu }$ and $g_{\lambda \mu \nu}$ are CPT-odd, while $d_{\mu
\nu}$ and $H_{\mu \nu }$ are CPT-even fields. Five and
six-dimensional operators can also lead to spin coupling terms
\cite{Posp3,Posp4}. To leading order the spin energy shift can be
written as
\begin{equation}
\delta E = -\mu_{^{3} \rm He} \beta_i^N \sigma_i^N =- P_n
\widetilde{b}_i^n \sigma_i^n -2 P_p \widetilde{b}_i^p \sigma_i^p
\end{equation}
where $P_n=0.87$ and $P_p=-0.027$ are the polarizations of neutron
and proton in the $^3$He nucleus \cite{Friar} and $\sigma_i$ are the
Cartesian components of the spin Pauli matrix. $\widetilde{b}$ is
defined in terms of coefficients in Eq. (\ref{Lagr})  in Ref.
\cite{clockcom}.

Taking only the leading order neutron spin coupling, we obtain
\begin{eqnarray}
\widetilde{b}^n_X &= &(0.1 \pm 1.6) \times 10^{-33}~{\rm GeV} \\
\widetilde{b}^n_Y &= &(2.5 \pm 1.6) \times 10^{-33}~{\rm GeV}
\end{eqnarray}
which can be interpreted as $|\widetilde{b}^n_{XY}|<3.7 \times
10^{-33}$ GeV at 68\% confidence level. Our measurement is also
sensitive to proton coupling $\widetilde{b}^p_{XY}$ at a level of $6
\times 10^{-32}$ GeV and can be used in combination with the
$^3$He-$^{129}$Xe maser result \cite{Bear} and recent analysis of
the nuclear spin content of $^{129}$Xe \cite{Flambaum} to set an
independent stringent limit on proton Lorentz violation. The limits
on models involving higher dimension operators
\cite{Posp1,Posp2,Posp3} are also improved. The limits on
boost-dependent Lorentz and CPT violation effects \cite{cane} can be
improved from comparison of winter and summer data. Even though we
measure only spatial components of spin anisotropy in the Earth's
equatorial plane, time-like components of Lorentz violation could
also be observed since the solar system is moving relative to the
rest frame of the cosmic microwave background (CMB) with a velocity
$v \sim 10^{-3}c$ at a declination of $-7\degree$. Possible
anisotropy associated with the alignment of low-order CMB multipoles
also points in approximately the same direction \cite{Scwartz}.

In conclusion, we have set a new limit on neutron spin coupling to a
Lorentz and CPT violating background field. Our results represent a
factor of 30 improvement of the previous limit for the neutron and
have the highest energy resolution of any spin anisotropy
experiment. The fundamental limits of the co-magnetometer
sensitivity have not yet been realized. For example, in a stationary
co-magnetometer we have achieved energy sensitivity of $10^{-34}$
GeV \cite{spinforce}. Additional improvements by 1-2 orders of
magnitude are expected from the use of $^{21}$Ne in the
co-magnetometer \cite{Ghosh}. The main systematic effects are due to
a combination of coupling to Earth's rotation and gravity. They can
be reduced by placing the experiment near the South Pole to avoid
relying on a sidereal variation to extract the Lorentz-violation
signal. With these improvements it should be possible to achieve
energy sensitivity on the order of $10^{-36}$ GeV, reaching the
level needed to observe effects suppressed by two powers of the
Plank mass, such as dimension-6 Lorentz-violating operators
\cite{Posp4}.

This work was supported by NSF Grant No. PHY-0653433 and DARPA.

\end{document}